\documentstyle[aas2pp4,epsf]{article}

\lefthead{Potekhin, Varshalovich, Lanzetta et al.}
\righthead{Cosmological Variability of the Proton Mass}

\begin{document}

\title{TESTING COSMOLOGICAL VARIABILITY OF THE PROTON-TO-ELECTRON MASS RATIO 
USING THE SPECTRUM OF PKS 0528--250 }
\author{Alexander Y. Potekhin, Alexander V. Ivanchik, and
Dmitry A.~Varshalovich}
\affil{Ioffe Physical-Technical Institute, 
194021 St.-Petersburg, Russia}
\author{Kenneth M. Lanzetta}
\affil{Astronomy Program, Department of Earth and Space Sciences \\
State University of New York at Stony Brook, Stony Brook, NY
11794--2100, U.S.A.} 
\author{Jack A. Baldwin}
\affil
{Cerro Tololo Inter-American Observatory, Casilla 603, La Serena,
Chile}
\author{Gerard M. Williger}
\affil{Goddard Space Flight Center,
Code 681, NASA/GSFC,
Greenbelt MD 20771,
U.S.A.}
\and 
\author{R.~F. Carswell}
\affil{Institute of Astronomy, Madingley Road, Cambridge CB3 0HA, U.K.} 

\begin{abstract}

  Multidimensional cosmologies allow for variations of fundamental physical
constants over the course of cosmological evolution, and different versions 
of the theories predict different time dependences.  In particular, such
variations could manifest themselves as changes of the proton-to-electron mass
ratio $\mu=m_{\rm p}/m_{\rm e}$ over the period of $\sim 10^{10}$ yr since the
moment of formation of high-redshift QSO spectra.  Here we analyze a new,
high-resolution spectrum of the $z = 2.81080$ 
molecular hydrogen absorption system toward the QSO
PKS 0528--250 to derive a new observational constraint to the time-averaged
variation rate of the proton-to-electron mass ratio. We find $| \dot{\mu}/
\mu | < 1.5 \times 10^{-14}$ yr$^{-1}$, which is much tighter than previously
measured limits. 

\end{abstract}


\section{INTRODUCTION}

  The possibility of the variability of fundamental physical constants was
first put forward by Dirac (1937) in the course of his discussion with Milne
(1937).  Later it was considered by Teller (1948), Gamow (1967),
Dyson (1972) and other physicists.  Interest in the problem increased
greatly during the last decade, 
due to new developments in the Kaluza--Klein and
supergravity models of unification of all the physical interactions.  
Chodos \&
Detweiler (1980), Freund (1982), Marciano (1984), and Maeda (1988) discussed 
possibilities of including these multidimensional theories into the 
cosmological scenario of the expanding Universe and found that the low-energy
limits to the fundamental constants might vary over the cosmological
time. 
Variations of the coupling constants of strong and electroweak
interactions might then cause the masses of elementary particles to 
change. 
Note that an increase of the proton 
mass by 0.08\% would lead to transformation of protons into neutrons 
(by electron capture), resulting in destruction of atoms in the 
Universe. 
As demonstrated by Kolb, Perry, \& Walker (1987) 
and Barrow (1987), observational bounds on 
the time evolution of extra spatial dimensions in the Kaluza--Klein 
and superstring theories can be obtained 
from limits on possible variations of 
the coupling constants. 
Damour \& Polyakov (1994) have developed 
a modern version of the string theory which assumes
cosmological variations of the coupling constants and 
hadron-to-electron mass ratios. 
Therefore the parameters of the theory can be 
restricted by testing cosmological changes of 
these ratios. 

  The present value of the proton-to-electron mass ratio is
$\mu=1836.1526645\,(57)$ (CODATA, 1997).
Obviously, any significant 
variation of this parameter over a small time interval is excluded, but such
variation over the cosmological time $\sim 1.5\times 10^{10}$ yr remains a
possibility.  This possibility can be checked by analyzing spectra of
high-redshift QSOs.

The first analysis of this kind has been performed by Pagel (1977),
who obtained a restriction 
$|\dot{\mu}/\mu|< 5\times 10^{-11}{\rm ~yr}^{-1}$
on the variation rate of $\mu$ by comparison of 
wavelengths of H\,I and heavy-ion absorption lines, as proposed by 
Thompson (1975). This technique, however, could not provide 
a fully conclusive result, since the heavy elements 
and hydrogen usually belong to different interstellar clouds,
moving with different radial velocities.
In this paper we use another technique,
based on an analysis of H$_2$ absorption lines only. 

  One object suitable for such analysis is the $z = 2.811$ absorption system
toward PKS 0528--250, in which Levshakov \& Varshalovich (1985) identified
molecular hydrogen absorption lines based on a spectrum 
obtained by Morton et al.\ (1980).  Foltz, Chaffee, \& Black (1988) 
have presented a limit to possible variation of $\mu$
based on their observations of PKS 0528--250.  Their
analysis did not, however, take into account wavelength-to-mass sensitivity
coefficients, hence their result appeared to be not well grounded.
Subsequently the spectrum of Foltz, Chaffee, \& Black (1988) was reappraised by
Varshalovich \& Levshakov (1993), who obtained $|\Delta\mu/\mu| < 0.005$ at the
redshift $z=2.811$, and by Varshalovich \& Potekhin (1995), who obtained
$|\Delta\mu/\mu| < 0.002$ at the $2\sigma$ significance level.  (Here
$\Delta\mu/\mu$ is the fractional variation of $\mu$.)  
More recently, Cowie \& Songaila (1995) used a new spectrum of
PKS 0528$-$250 obtained with the Keck telescope to arrive at the 95\%
confidence interval $-5.5 \times 10^{-4} < \Delta\mu/\mu < 7 \times10^{-4}$. 

  Here we present a profile fitting analysis of a new, high-resolution
spectrum of PKS 0528$-$250, obtained in November 1991 with the Cerro-Tololo 
Inter-American Observatory (CTIO) 4 m telescope. We have calculated the 
wavelength-to-mass sensitivity coefficients for a larger number of spectral
lines and employed them in the analysis, which yields the strongest
observational constraint yet to possible $\mu$ variation over the cosmological
time scale (eq.~[\ref{limits}] below). 

\section{OBSERVATIONS}

Observations were obtained with the CTIO 4 m telescope in a series 
of exposures of typically 2700 s duration totaling 33750 s duration. 
The CTIO Echelle Spectrograph with the Air Schmidt camera and Reticon
CCD was used at the Cassegrain focus to obtain complete spectral coverage
over the wavelength range $\lambda \approx 3465-4905$ \AA. Observations of 
standard stars and of a Th-Ar comparison arc lamp were obtained at 
intervals throughout each night, and observations of a quartz lamp 
were obtained at the beginning or end of each night.  For all 
observations, the slit was aligned to the parallactic angle.

Data reduction was performed following procedures similar to 
those described previously by Lanzetta et al.\ (1991).  
One-dimensional spectra were extracted from the two-dimensional 
images using an optimal extraction technique, and individual 
one-dimensional spectra were coadded using an optimal coaddition 
technique.  Wavelength calibrations were determined from 
two-dimensional polynomial fits to spectral lines obtained in the 
Th-Ar comparison arc lamp observations. 
Continua were fitted to the one-dimensional 
spectra using an iterative spline fitting technique.

  The spectral resolution was measured from spectral lines obtained 
in the Th-Ar comparison arc lamp exposures.  This is appropriate 
because for all observations the seeing profile was larger than the 
slit width.  The spectral resolution was found to be  
${\rm FWHM} \approx 21-24$ km s$^{-1}$ 
in the spectral intervals used for the analysis. 

\begin{figure*}
\epsfysize=175mm
\epsffile[0 120 390 700]{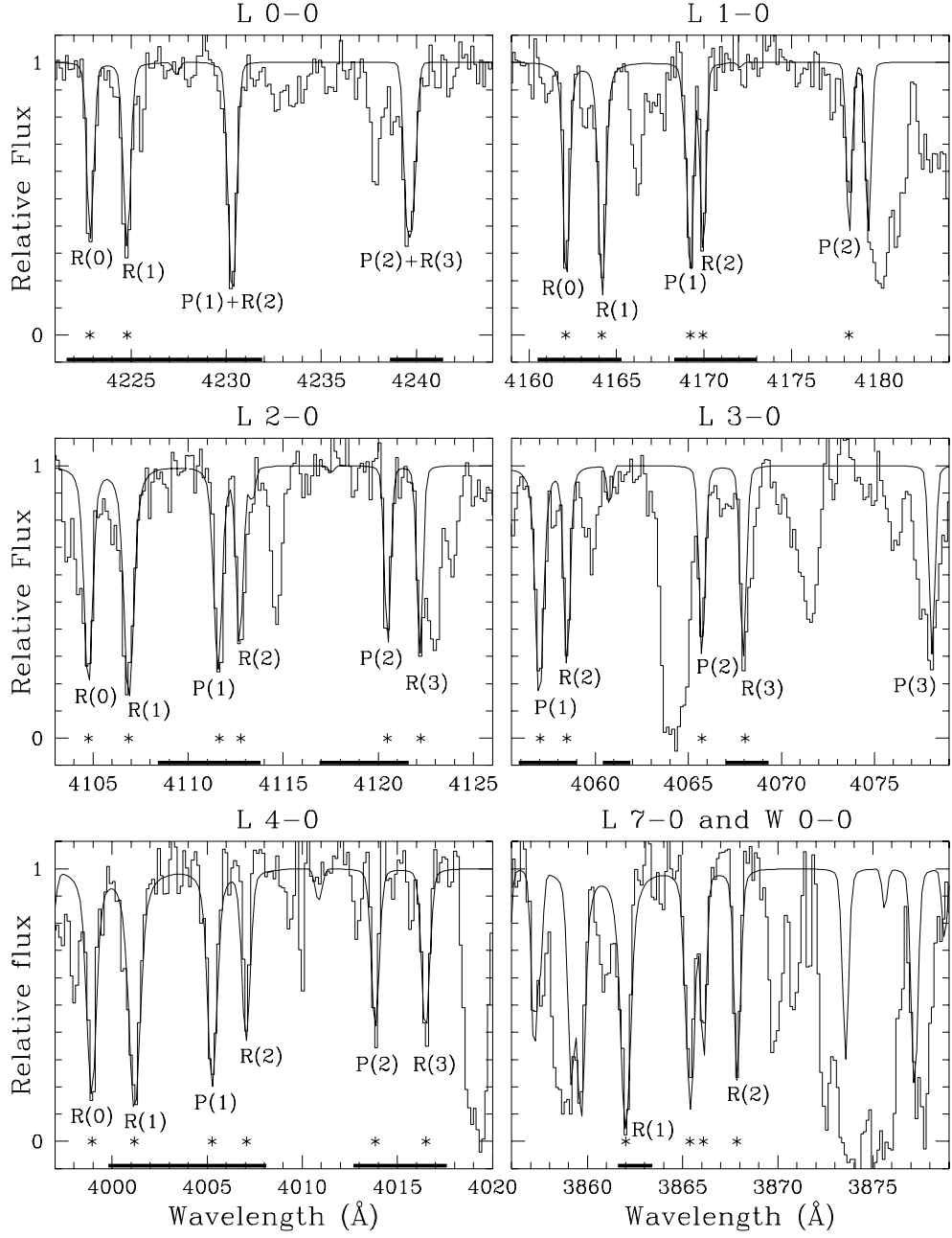}
\caption{Selected parts of the spectrum PKS 0528$-$250 
with absorption lines of H$_2$ molecules at the redshift 
$z=2.81080$. The fit is superposed on the data. 
The most distinct absorption lines of the Lyman band 
are labeled on the plot. 
The thick bars along the horizontal axes mark 
the spectral intervals used in the fit (Sect.~4.1),
and the asterisks mark the positions 
of individual lines listed in Table~\ref{tab2}
and used in the independent analysis in Section~4.2
(note that there are other spectral intervals and lines
included in the analyses but belonging to 
Lyman and Werner branches not shown in the figure).
}
\label{fig1}
\end{figure*}

Figure \ref{fig1} presents parts of the spectrum, with the fit superposed on 
the data, for several spectral intervals in which the H$_2$ absorption 
lines have been analyzed 
(for more detail, see {\'C}ircovi{\'c} et al., 1998). 

\section{SENSITIVITY COEFFICIENTS} 

The possibility 
of distinguishing between the cosmological redshift of the spectrum 
and wavelength shifts due to a variation of $\mu$ arises from the 
fact that the electronic, vibrational, and rotational energies of a 
molecule each undergo a different dependence on the reduced mass of 
the molecule.  Hence comparison of the wavelengths of various 
electronic-vibrational-rotational molecular absorption lines 
observed in the spectrum of a high-redshift quasar with 
corresponding molecular lines observed in the laboratory 
may reveal or limit the variation of $\mu$ with time.  

If the value of $\mu$ at the early epoch $z$ of the absorption spectrum 
formation were different 
from the contemporary one, then the ratio
\begin{equation}
{(\lambda_i/\lambda_k)_z \over (\lambda_i/\lambda_k)_0} \simeq 
1+(K_i-K_k)\left({\Delta\mu\over\mu}\right)
\label{1}
\end{equation}
would deviate from unity. Here 
\begin{equation}
K_i = {\rm d}\ln\lambda_i/{\rm d}\ln\mu
\label{02}
\end{equation}
is the coefficient which determines the sensitivity of the 
wavelength  $\lambda_i$ of $i$th spectral line with respect to the 
variation of the mass ratio $\mu$. 

These coefficients were calculated previously by Varshalovich 
\& Levshakov (1993) from the spectroscopic constants of the 
H$_2$ molecule, using the Born--Oppenheimer 
approximation. Later Varshalovich \& Potekhin (1995) 
calculated $K_i$ in another way, by comparison of the H$_2$ 
laboratory wavelengths with the corresponding wavelengths for D$_2$ 
and T$_2$ molecules, which simulate just the mass variation of the
study, and also for HD molecules. Varshalovich \& Potekhin (1995) 
also removed some inaccuracies from the table of
Varshalovich \& Levshakov (1993). The two ways of performing the calculation 
yielded very similar $K_i$ values, which argues that both are correct.

For each 
electron-vibration-rotational band, a wavelength of a transition 
between two states with the vibrational and rotational quantum 
numbers $v,~J$ and $v',~J'$ can be presented as
\begin{equation}
\lambda = 
\left[\nu^u_{v'J'}-\nu^l_{v''J''}\right]^{-1}, 
\label{03}
\end{equation}
where 
$\nu$ is the level energy in cm$^{-1}$, and 
the superscripts $u$ and $l$ stand for the upper and the 
lower level, respectively. For each of them 
\begin{equation}
\nu_{vJ} = \sum_{m,n} 
Y_{mn}
\left(v+{1\over 2}\right)^m 
(J(J+1))^n.
\label{nu}
\end{equation}
We consider the Lyman bands (transitions 
${\rm X}\,^1\Sigma_g^+\to{\rm B}\,^1\Sigma_u^+$) 
and the Werner bands 
(${\rm X}\,^1\Sigma_g^+\to{\rm C}\,^1\Pi_u^+$) of the molecular H$_2$ 
spectrum. The parameters $Y_{mn}$ for the three 
corresponding states are taken from Huber \& Herzberg (1979). 
The coefficient $Y_{00}$ is redefined so that the energy 
of each rotational-vibrational band is counted from 
the ground-state energy. 
For the 
state $\Pi_u$, the factor $J(J+1)$ in the terms with $n=1$ 
of equation (\ref{nu}) has been replaced by $(J(J+1)-1)$, 
in order to take into account the projection ($\Lambda^2=1$) 
of the electron orbital moment on the molecular axis. 

  From the Born--Oppenheimer approximation we conclude that the 
coefficients $Y_{mn}$ are proportional to $\mu^{-n-m/2}$. 
Then the 
sensitivity coefficients $K_i$ are easily obtained from 
equations (\ref{02})--(\ref{nu}):
\begin{equation}
   K_{v'J'-v''J''} = \lambda_{v'J'-v''J''} 
   \left(k^u_{v'J'} - k^l_{v''J''}\right),
\label{K}
\end{equation}
where 
\begin{equation}
   k_{vJ} = \sum_{m,n} y_{mn}
\left(v+{1\over 2}\right)^m 
(J(J+1))^n,  
\label{k}
\end{equation}
and the coefficients $y_{mn}$ are given in Table~\ref{tab1} (in cm$^{-1}$). 
For the state $\Pi_u$, the factor $J(J+1)$ in the terms with $n=1$ 
has been replaced by $(J(J+1)-1)$, as well as in equation (\ref{nu}).

\pagebreak
\section{RESULTS OF ANALYSIS}

\subsection{Synthetic spectrum analysis}

We have applied to the spectrum a 
routine described previously by Lanzetta \& Bowen (1992). 
This routine 
performs a comparison of the synthetic and observed spectrum and
finds an optimal solution to a parameterized model of a 
set of absorption profiles, simultaneously taking account of all 
observed spectral regions and transitions. Parameter estimates are 
determined by minimizing $\chi^2$, 
and parameter uncertainties and correlations are determined by 
calculating the parameter covariance matrix at the resulting 
minimum. 

A total of 59 H$_2$ transitions are incorporated into the $\chi^2$ 
fit, and the absorption lines corresponding to these transitions 
occur across the linear, saturated, and damped parts of the curve of 
growth. The redshift, Doppler parameter,
and column densities of the H$_2$ rotational levels $J'' = 0$ through $J'' =
7$ were adopted as free parameters.  Wavelengths, oscillator strengths, and
damping coefficients of the H$_2$ transitions were taken from Morton \&
Dinerstein (1976). 
According to {\'C}ircovi{\'c} et al.\ (1998), 
the total H$_2$ column density is 
$
\log N({\rm H}_2) = 18.45\pm 0.02 
$
and the Doppler parameter is
$
b = 3.23 \pm 0.11
$
at the redshift 
\begin{equation}
  z=2.8107998\,(24).
\label{z}
\end{equation}
Full details of the reduction and analysis of the spectrum 
are described in a companion paper ({\'C}ircovi{\'c} et al., 1998), 
including the list of 
all spectral intervals and transitions used in the fit. 
These spectral intervals (shown in Fig.~\ref{fig1} by horizontal bars)
were chosen to embrace anticipated positions 
of distinct and presumably unblended H$_2$ lines.
We emphasize that, 
although the choice of the window function is somewhat arbitrary,
it should not entail systematic shifts of the parameter estimates.
Note that there are H$_2$ lines present 
in the spectrum but not used in the fit. 
Some of them are seen in Figure~\ref{fig1} (for example, 
L~1--0 P(2), L~7--0 P(1) and R(2), and others). 
The wavelengths and strengths of these lines
are perfectly reproduced by the synthetic spectrum.
Furthermore, within the errors, 
none of the model lines drops below the observed spectrum.
This remarkable agreement between the measured and model synthetic spectrum 
confirms the reliability of the derived parameters. 

\begin{figure}[t]
\epsfysize=55mm
\epsffile[80 290 410 560]{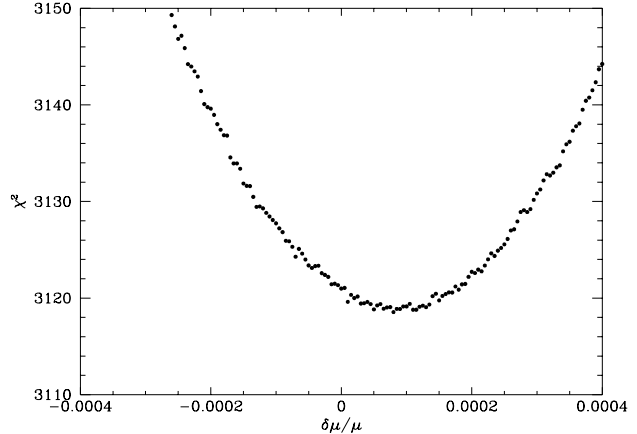}
\caption{Best fit (with respect to all other parameters of the problem)
dependence of $\chi^2$ on \ $\Delta\mu/\mu$.  
There are 1367 degrees of freedom
in the $\chi^2$ fitting analysis.}
\label{fig2}
\end{figure}

  A limit to the variation of the proton-to-electron inertial mass ratio 
was obtained
by repeating the $\chi^2$ synthetic spectrum fitting analysis 
with an additional free parameter $\Delta\mu/\mu$.  
The dependence of $\chi^2$ on this parameter is shown 
in Figure \ref{fig2}. 
The resulting parameter estimate and $1\sigma$ uncertainty is
\begin{equation}
\Delta \mu/\mu = (8.3^{+6.6}_{-5.0})\times 10^{-5}. 
\label{dmu}
\end{equation}
This result\footnote{
The estimate (\ref{dmu}) has been presented 
at the 17th Texas Symp.\ on Relativistic Astrophys. 
(Varshalovich et al., 1996). 
} 
indicates a value of $\Delta \mu / \mu$
that differs from $0$ at the $1.6 \sigma$ level. 
The $2 \sigma$ confidence interval to $\Delta \mu / \mu$ is 
\begin{equation}
-1.7\times 10^{-5} < \Delta \mu / \mu < 2\times 10^{-4}. 
\label{limits}
\end{equation}
Assuming that the age of the Universe is $\sim 15$ Gyr 
the redshift $z=2.81080$ corresponds to the elapsed time of 13 Gyr 
(in the standard cosmological model with $\Omega=1$). 
Therefore we arrive at the restriction 
\begin{equation}
|\dot{\mu}/\mu|< 1.5\times 10^{-14}{\rm ~yr}^{-1}
\label{constraint}
\end{equation}
on the variation rate of $\mu$, averaged over 87\% of 
the lifetime of the Universe. 

\subsection{Profile analysis of separate lines}

We have analyzed the spectrum also by another, traditional, technique.
The use of the alternative technique provides an independent check 
for the results of the above $\chi^2$ analysis and enables a direct 
comparison with the previous results (Foltz, Chaffee, \& Black, 1988; 
Varshalovich \& Levshakov, 1993; Varshalovich \& Potekhin, 1995; 
Cowie \& Songaila, 1995). 

We have selected spectral lines of the
Lyman and Werner bands which 
can be unambiguously fitted by a single Gaussian profile
and a few lines whose decomposition in 
two contours is quite certain (such as L~1--0 P(1) and R(2)
at $\lambda_z=4170$\,\AA\ and L~7--0 P(1) and W~0--0 P(3) 
at $\lambda_z=3866$\,\AA).
Since there are overlapping diffraction orders,
we have selected them so as to work far from the order edges,
and therefore the resolution in the analyzed regions was
relatively high ($R>10\,000$) and uniform.
The analyzed 50 transitions
are listed in the first column of Table~\ref{tab2} 
and marked in Figure~\ref{fig1} by asterisks.
Only 26 of these 50 lines have been included in the analysis 
of the synthetic spectrum described in Section 4.1,
so that the total number of H$_2$ wavelengths analyzed by both techniques 
amounts to 83. 
Thus in this section we use not only an independent technique 
but also an independent choice of the spectral regions.

\begin{figure}[h]
\epsfysize=55mm
\epsffile[85 290 525 585]{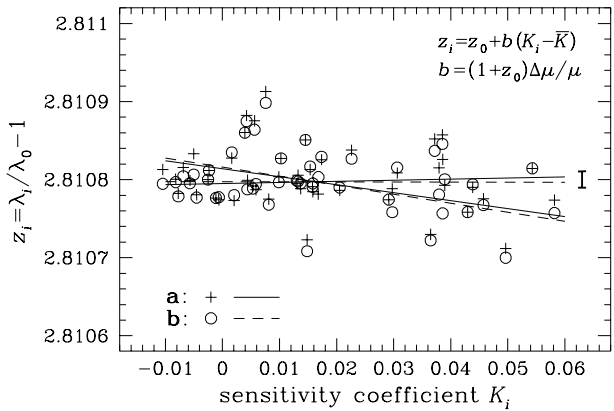}
\caption{Relative deviations of the redshift values, 
inferred from an analysis of separate spectral features, 
plotted vs.\ sensitivity coefficients.
The lines represent
$2\sigma$ deviations from the slope $b$
of the best linear regression.
The results
based on the rest-frame wavelengths by 
({\bf a}) Abgrall et al.\ (1993) 
(crosses and solid lines) and ({\bf b}) 
Morton \& Dinerstein (1976)
(circles and dashes) are shown.
The errorbar to the right represents the $\pm2\sigma$ 
limit on $z_0$.}
\label{fig3}
\end{figure}

The rest-frame wavelengths $\lambda_0$ 
adopted from Abgrall et al.\ (1993)
are given in the second column of Table~\ref{tab2}.
The third and fourth columns of Table~\ref{tab2} present 
the optimal vacuum heliocentric position of the center 
of each observed profile ($\lambda_z$) and the 
estimated standard deviation ($\sigma_\lambda$). 
The values of $\lambda_z$ and $\sigma_\lambda$
have been provided simultaneously by the standard fitting
procedure that minimized root-mean-square deviations
between the fit and data. 
Sensitivity coefficients $K_i$, calculated according to 
equations (\ref{K}), (\ref{k}), are listed in the fifth column. 
The last column presents the redshift corresponding to each $\lambda_z$.
These redshifts, $z_i$, are  shown by crosses in Figure~\ref{fig3}.
In order to test the influence of possible uncertainty in 
$\lambda_0$,
we have repeated the analysis using a set of the wavelengths 
by Morton \& Dinerstein (1976); the corresponding redshifts 
are shown in Figure~\ref{fig3} with open circles.

In the linear approximation, 
$
z(K_i)=z+b K_i,
$
where $b=(1+z)\Delta\mu/\mu$, and $z$ is the cosmological redshift
of the H$_2$ system. In order the estimates of the regression 
parameters to be statistically independent, it is convenient to 
write the regression in the form
\begin{equation}
   z_i=z_0+b (K_i-\bar{K}),
\label{regression}
\end{equation}
where $\bar{K}$ is the mean value of $K_i$.

Given the typical values  of $\sigma_\lambda\sim0.02$\,\AA\ and
$\lambda\sim4000$\,\AA, one has a typical
relative error 
$\sigma_z\approx(\sigma_\lambda/\lambda)(1+z)\sim2\times10^{-5}$
for an individual line. 
This estimate is only an intrinsic statistical error,
and it does not include an error
due to possible unresolved blends. 
For this reason, we have not relied on the estimated $\sigma_\lambda$
values in our statistical analysis but
calculated the $1\sigma$ uncertainties
from the actual scatter of the data
(see also a discussion in Potekhin \& Varshalovich, 1994).

The estimated mean and slope parameters 
of the linear regression (\ref{regression}),
based on $\lambda_0$ by Morton \& Dinerstein (1976)
are $z_0=2.8107973\,(52)$
and
$
   b=(-5.85\pm2.86)\times10^{-4}.
$
Using the data of Abgrall et al.\ (1993),
we obtain $z_0=2.8108028\,(53)$
and
$
   b=(-4.38\pm2.91)\times10^{-4}.
$
The dashed and solid lines in Figure~\ref{fig3}
correspond to the $2\sigma$-deviations of $b$ 
for the first and second set of $\lambda_0$, respectively.
The errorbar to the right represents the $2\sigma$ 
limit on $z_0$.

The latter estimate of $b$ translates into 
\begin{equation}
   \Delta\mu/\mu = (-11.5\pm 7.6)\times 10^{-5}.
\label{dmu2}
\end{equation}
When using the weights $\propto\sigma_\lambda^{-2}$,
we obtain a similar estimate, 
$
   \Delta\mu/\mu = (-10.2\pm 8.1)\times 10^{-5}.
$
Since the distribution of random errors caused 
by different sources (including possible blends)
is not expected to be Gaussian, it may worth
using robust statistical techniques such as
the trimmed-mean regression 
analysis (cf. Potekhin \& Varshalovich, 1994). 
We have applied the trimmed-mean technique of Ruppert \& Carroll (1980)
and found that, 
at any trimming level up to 12\%, the estimate of $b$
is closer to zero but has a larger estimated $1\sigma$
error, compared with 
the result of the standard least-square analysis
reported in equation~({\ref{dmu2}). 
Thus we adopt equation~({\ref{dmu2}) 
as the final result of this section. 

The estimate (\ref{dmu2}) has a larger 
statistical error compared with equation (\ref{dmu}). 
Within $2\sigma$, both estimates are consistent 
with the null hypothesis of no variation of $\mu$. 

\section{CONCLUSIONS}

We have obtained a constraint to the variation rate of 
the proton-to-electron mass ratio $\mu$. 
Two fitting procedures have been used, 
one of which simultaneously takes into
account all observed spectral regions and transitions, while the other 
is applied to each spectral feature separately. 
The two techniques, applied to two different sets 
of spectral intervals, have resulted in similar 
upper bounds on $\Delta\mu/\mu$, at the level $\,\sim2\times10^{-4}$. 
The obtained restriction on $\dot{\mu}/\mu$ (\ref{constraint}) 
is by an order of magnitude 
more stringent than the limit set previously by 
Varshalovich \& Potekhin (1995), who used a spectrum with a lower 
spectral resolution. Moreover, it is much 
more restrictive than the estimate of Cowie \& 
Songaila (1995), based on high-resolution Keck telescope 
observations. 
There are two reasons for the higher accuracy of the
present estimate. 
First, our fitting procedure simultaneously takes into
account all observed spectral regions and transitions.  This is
particularly important because many of the transitions are blended, even at
the spectral resolution of the spectrum used by Cowie \& Songaila (1995). 
A separate analysis of spectral lines leads to larger statistical 
errors, as we have shown explicitly in Section 4.2. 
Second, we include a larger number of transitions 
between excited states of the H$_2$ molecule
(83 spectral lines, compared with 19 lines used by Cowie \& Songaila),
many of which have higher wavelength-to-mass sensitivity coefficients $K_i$. 
The larger interval of $K_i$ values results in a higher 
sensitivity to possible mass ratio deviations.

The method used here to determine 
the variation rate of $\mu$ 
could be formally less sensitive 
than the one based on an analysis of 
relative abundances of chemical elements produced 
in the primordial nucleosynthesis (Kolb et al., 1986).  
However, the latter method is very indirect
since it depends on a physical model 
which includes a number of additional assumptions. 
Therefore the present method seems to be more reliable.

Quite recently, Wiklind \& Combes (1997)
used a similar method (following Varshalovich \& Potekhin, 1996)
in order to infer limits on time variability
of masses of molecules CO, HCN, HNC and 
the molecular ion HCO$^+$ from high-resolution
radio observations of rotational 
lines in spectra of a few low-redshift ($z<1$) quasars.
The result reported in this paper 
constrains the mass of the H$_2$ molecule,
and thus the proton mass, at much larger $z$.
These constraints
may be used for checking the 
multidimensional cosmological models which predict time-dependences 
of fundamental physical constants. The described method of the 
calculation of the sensitivity coefficients can also be used for 
analyzing any other high-redshift molecular clouds, which may be 
found in future observations. 

\begin{acknowledgements}
AYP, AVI and DAV acknowledge partial
support from RBRF grant 96-02-16849 and ISF grant NUO\,300.
KML was supported by NASA grant NAGW--4433 
and by a Career Development Award
from the Dudley Observatory.
\end{acknowledgements}

\begin{table}[t]
\caption{Coefficients $y_{mn}$, in cm$^{-1}$}
\label{tab1}
\begin{tabular}{lccc}
\noalign{\smallskip}
\hline
\hline
\noalign{\smallskip}
 &X$^1\Sigma_g^+$ & B$^1\Sigma_u^+$ & C$^1\Pi_u$ \\
\noalign{\smallskip}
\hline
\noalign{\smallskip}
$y_{10}$  &  2200.607          &   679.05   &  1221.89  \\
$y_{20}$  &$-$121.336          &$-$20.888   &$-$69.524  \\
$y_{30}$  &  1.2194            &  1.0794    &  1.0968   \\
$y_{40}$  &                    &$-$0.1196   &$-$0.0830  \\
$y_{50}$  &                    &  0.00540   &           \\
$y_{01}$  &  60.8530           &  20.01541  &  31.3629  \\
$y_{11}$  &$-$4.0622           &$-$1.7768   &$-$2.4971  \\
$y_{21}$  &  0.1154            &  0.2428    &  0.0592   \\
$y_{31}$  &$-$0.0128           &$-$0.0293   &$-$0.00740 \\
$y_{41}$  &                    &  0.00138   &           \\
$y_{02}$  & $-$0.0942          & $-$0.03250 & $-$0.0446 \\
$y_{12}$  &0.00685          &0.005413 &0.00185 \\
$y_{22}$  &  $-$0.0012         & $-$0.0006867 &         \\
$y_{32}$  &        &$4.148\times 10^{-5}$ &             \\
$y_{03}$  &$1.38\times 10^{-4}$&            &           \\
\noalign{\smallskip}
\hline
\noalign{\smallskip}
\end{tabular}
\end{table}

\begin{table*}[h]
\caption{H$_2$ lines and sensitivity coefficients}
\label{tab2}
\small
\begin{tabular}{lrrcrr}
\noalign{\smallskip}
\hline
\hline
\noalign{\smallskip}
 Line & $\lambda_0$ (\AA)~ & $\lambda_z$ (\AA)~
& $\sigma_\lambda$ (\AA) & $K_\lambda$~~~ & $z_\lambda$~~~~ \\
\noalign{\smallskip}
\hline
\noalign{\smallskip}
L 0--0 R(1) & 1108.633 & 4224.779 & 0.014 & $-0.00818$ & 2.8107969 \\
L 0--0 R(0) & 1108.127 & 4222.830 & 0.010 & $-0.00772$ & 2.8107782 \\
L 1--0 P(2) & 1096.438 & 4178.284 & 0.012 & $-0.00453$ & 2.8107765 \\
L 1--0 R(2) & 1094.244 & 4169.945 & 0.012 & $-0.00252$ & 2.8108000 \\
L 1--0 P(1) & 1094.052 & 4169.226 & 0.012 & $-0.00234$ & 2.8108116 \\
L 1--0 R(1) & 1092.732 & 4164.157 & 0.008 & $-0.00113$ & 2.8107761 \\
L 1--0 R(0) & 1092.195 & 4162.108 & 0.007 & $-0.00064$ & 2.8107772 \\
L 2--0 R(3) & 1081.712 & 4122.218 & 0.011 & $ 0.00165$ & 2.8108347 \\
L 2--0 P(2) & 1081.267 & 4120.463 & 0.015 & $ 0.00206$ & 2.8107800 \\
L 2--0 R(2) & 1079.226 & 4112.779 & 0.012 & $ 0.00394$ & 2.8108598 \\
L 2--0 P(1) & 1078.923 & 4111.648 & 0.008 & $ 0.00422$ & 2.8108747 \\
L 2--0 R(1) & 1077.697 & 4106.879 & 0.016 & $ 0.00535$ & 2.8107884 \\
L 2--0 R(0) & 1077.140 & 4104.751 & 0.080 & $ 0.00587$ & 2.8107940 \\
L 3--0 R(3) & 1067.474 & 4068.050 & 0.095 & $ 0.00758$ & 2.8108982 \\
L 3--0 P(2) & 1066.899 & 4065.712 & 0.117 & $ 0.00812$ & 2.8107678 \\
L 3--0 R(2) & 1064.993 & 4058.479 & 0.020 & $ 0.00989$ & 2.8107963 \\
L 3--0 P(1) & 1064.606 & 4057.029 & 0.019 & $ 0.01026$ & 2.8108267 \\
L 4--0 R(3) & 1053.977 & 4016.490 & 0.012 & $ 0.01304$ & 2.8107983 \\
L 4--0 P(2) & 1053.283 & 4013.838 & 0.010 & $ 0.01369$ & 2.8107950 \\
L 4--0 R(2) & 1051.498 & 4007.062 & 0.010 & $ 0.01536$ & 2.8108164 \\
L 4--0 P(1) & 1051.031 & 4005.259 & 0.008 & $ 0.01580$ & 2.8107905 \\
L 4--0 R(1) & 1049.964 & 4001.183 & 0.030 & $ 0.01681$ & 2.8108029 \\
L 4--0 R(0) & 1049.367 & 3998.954 & 0.014 & $ 0.01736$ & 2.8108286 \\
L 5--0 R(4) & 1044.542 & 3980.460 & 0.031 & $ 0.01485$ & 2.8107082 \\
L 5--0 P(3) & 1043.501 & 3976.557 & 0.005 & $ 0.01584$ & 2.8107950 \\
L 6--0 P(3) & 1031.192 & 3929.652 & 0.050 & $ 0.02053$ & 2.8107897 \\
L 6--0 R(3) & 1028.983 & 3921.287 & 0.010 & $ 0.02262$ & 2.8108264 \\
L 7--0 R(2) & 1014.977 & 3867.848 & 0.007 & $ 0.02914$ & 2.8107740 \\
W 0--0 P(3) & 1014.504 & 3866.085 & 0.014 & $-0.01045$ & 2.8107942 \\
L 7--0 P(1) & 1014.326 & 3865.381 & 0.009 & $ 0.02976$ & 2.8107576 \\
L 7--0 R(1) & 1013.436 & 3862.010 & 0.015 & $ 0.03062$ & 2.8108155 \\
W 0--0 Q(2) & 1010.938 & 3852.498 & 0.010 & $-0.00686$ & 2.8108040 \\
W 0--0 Q(1) & 1009.771 & 3848.034 & 0.011 & $-0.00570$ & 2.8107949 \\
W 0--0 R(2) & 1009.023 & 3845.218 & 0.014 & $-0.00503$ & 2.8108064 \\
L 9--0 R(2)  & 993.547 & 3786.139 & 0.032 & $ 0.03647$ & 2.8107220 \\
L 9--0 P(1)  & 992.809 & 3783.448 & 0.023 & $ 0.03719$ & 2.8108365 \\
L 9--0 R(1)  & 992.013 & 3780.378 & 0.014 & $ 0.03796$ & 2.8107804 \\
L 9--0 R(0)  & 991.376 & 3777.961 & 0.042 & $ 0.03858$ & 2.8107564 \\
W 1--0 R(3)  & 987.447 & 3762.962 & 0.019 & $ 0.00439$ & 2.8107874 \\
W 1--0 R(2)  & 986.243 & 3758.449 & 0.018 & $ 0.00562$ & 2.8108636 \\
L 10--0 P(2) & 984.863 & 3753.172 & 0.009 & $ 0.03854$ & 2.8108453 \\
L 11--0 P(3) & 978.218 & 3727.786 & 0.011 & $ 0.03896$ & 2.8108005 \\
L 11--0 R(2) & 974.156 & 3712.273 & 0.013 & $ 0.04295$ & 2.8107582 \\
L 12--0 R(3) & 967.675 & 3687.606 & 0.029 & $ 0.04386$ & 2.8107937 \\
W 2--0 R(3)  & 966.778 & 3684.203 & 0.013 & $ 0.01324$ & 2.8107977 \\
W 2--0 R(2)  & 965.793 & 3680.493 & 0.009 & $ 0.01456$ & 2.8108508 \\
L 13--0 P(3) & 960.450 & 3660.059 & 0.017 & $ 0.04574$ & 2.8107672 \\
L 13--0 R(2) & 956.578 & 3645.243 & 0.010 & $ 0.04963$ & 2.8106998 \\
L 15--0 P(3) & 944.331 & 3598.670 & 0.042 & $ 0.05430$ & 2.8108142 \\
L 15--0 R(2) & 940.623 & 3584.501 & 0.029 & $ 0.05816$ & 2.8107571 \\
\noalign{\smallskip}
\hline
\noalign{\smallskip}
\end{tabular}
\end{table*}
\normalsize

\end{document}